\documentclass[aps,pra,twocolumn,superscriptaddress]{revtex4}
\usepackage{filecontents}

\usepackage{amssymb}
\usepackage{graphicx}
\usepackage{amsmath}
\usepackage{xcolor}

\newcommand{\be}{\begin{equation}}
\newcommand{\ee}{\end{equation}}
\newcommand{\ba}{\begin{eqnarray}}
\newcommand{\ea}{\end{eqnarray}}

\newcommand{\de}{\delta}
\newcommand{\eps}{\epsilon}
\newcommand{\la}{\lambda}
\newcommand{\si}{\sigma}
\newcommand{\Si}{\Sigma}
\DeclareMathOperator{\sgn}{sgn}
\newcommand\id{\mathbb{I}}
\newcommand\ket[1]{\left|#1\right\rangle}
\newcommand\vev[1]{{\left\langle{#1}\right\rangle}}
\newcommand\bbZ{\mathbb{Z}} 
\newcommand{\tc}{{t_c}}
\newcommand{\tL}{{t_L}}
\newcommand\dni{\de_{N\infty}}
\newcommand\Sni{\Si_{N\infty}}
\newcommand\epsbar{{\eps_0}}
\newcommand\dezero{{\de_0}}

\begin{document}

\preprint{UdeM-GPP-TH-16-253}

\title{Probing Anderson Localization using the Dynamics of a Qubit}
\author{Hichem Eleuch}
\email{heleuch@fulbrightmail.org}
\affiliation{Institute for Quantum Science and Engineering,
Texas A$\&$M University, College Station, Texas 77843, USA}
\affiliation{Department of Physics, McGill University, Montr\'eal, QC, Canada, H3A 2T8}
\affiliation{Groupe de physique des particules, Universit\'{e} de Montr\'{e}al, C.~P.~6128, Succursale Centre-ville, Montr\'eal, QC, Canada, H3C 3J7}
\author{Michael Hilke}
\email{hilke@physics.mcgill.ca}
\affiliation{Department of Physics, McGill University, Montr\'eal, QC, Canada, H3A 2T8}
\author{Richard MacKenzie}
\email{richard.mackenzie@umontreal.ca}
\affiliation{Groupe de physique des particules, Universit\'{e} de Montr\'{e}al, C.~P.~6128, Succursale Centre-ville, Montr\'eal, QC, Canada, H3C 3J7}

\begin{abstract}
Anderson localization is a consequence of coherent interference of multiple scattering events in the presence of disorder, which leads to an exponential suppression of the transmission. The decay of the transmission is typically probed at a given energy or frequency. Here we show that this decay affects the dynamics of a qubit coupled to the disordered system and we express the relaxation rate of the qubit in terms of the localization properties. Conversely, adding static disorder to a channel coupled to a qubit will reduce the decoherence rate of the qubit. Hence, when designing electrodes that couple to qubits it is possible to improve their performance by adding impurities to the channel.
\end{abstract}


\maketitle

\centerline{PACS: 73.20.Fz, 03.65.Yz, 73.21.La}

\section{Introduction}\label{sec-intro}

The dynamics of simple quantum systems has emerged as a powerful tool not only in the context of quantum information processing \cite{scully}, but more generally in new quantum technologies \cite{1a,1b}, often refereed to as the second quantum revolution \cite{dowling2003quantum}. Both rely on coherent oscillations between different quantum states. The decay of these oscillations is a sign of decoherence, either through relaxation, pure dephasing or a combination of both. The decoherence rate depends on the coupling to the environment. In solid-state-based qubits, this environment can be electrical (for instance, electrical leads or gates), magnetic (spins or magnetic fluctuations), or vibrational (phonons). In cold atoms, the decoherence environment is also given by optical absorption and electromagnetic fluctuations. A great deal of recent theoretical and experimental activity has focused on various approaches to decreasing the decoherence rate \cite{dec1,dec2,dec3,dec4}.

In the pioneering work of Legett and co-workers \cite{leggett1987dynamics}, the authors demonstrated the crucial role of the environment in the physics of small quantum systems. It was shown that the coupling of a two-level system (TLS) to an open system of either bosons (such as phonons) or electrons could effectively suppress the tunneling of the TLS. The spin-boson model is commonly studied in this context \cite{weiss1999quantum}, and the environment can be described by a spectral function with a power law and an exponential cutoff. For more structured environments,  less is known about the decoherence rate \cite{wilhelm2004spin}. The effect of intrabath interactions was presented in \cite{lages2005decoherence}, where the authors studied the decoherence of two coupled spins coupled to the random transverse Ising model. They found different decoherence rates depending on the nature of the spin-bath spectrum.

In general, the decoherence dynamics of a single spin attached to a bath has a long history \cite{cucchietti2005decoherence}. Disordered spin couplings have also been considered \cite{camalet2007effect}, as has the relationship between level statistics of the bath and the decoherence properties of the system \cite{brox2012importance}. More recently, the coupling of a single spin as a probe of many-body localization has also been considered \cite{PhysRevB.91.140202,van2016single}. In other spin systems, the transmission fidelity of a quantum state through a disordered spin chain has been studied \cite{zwick2012spin,Ronke2016}, as has quantum communication beyond the localization length in disordered spin chains using repeated quantum error corrections \cite{allcock2009quantum}.

Other systems have been considered as well, including ion traps \cite{1367-2630-12-12-123016} and mesoscopic detectors, where it was found that the conductance is sensitive to the state of a TLS \cite{pilgram2002efficiency}. Here the relaxation rate was found to be proportional to the resistance of the measuring device and vanishes at zero temperature due to the absence of thermal fluctuations \cite{makhlin2001quantum}. The dependence of the decoherence rate on other parameters such as the gate voltage have also been measured in experimental double-dot systems in isotopically pure silicon \cite{veldhorst2015two}.

\begin{figure}[h!]
	\centering
	\hspace*{0cm}\includegraphics[width=0.5\textwidth]{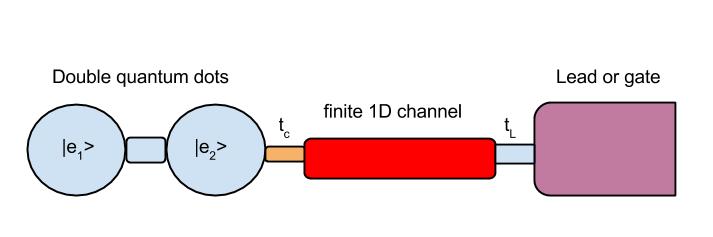}
	\caption{(color online) The TLS attached to a lead or a gate.}
	\label{sketch}
\end{figure}

In this work, we consider a model of a double quantum dot (TLS) first coupled directly to a semi-infinite chain (lead) and then coupled to such a chain via a one-dimensional channel of finite length, as illustrated in Fig.~\ref{sketch}. The semi-infinite chain is a periodic system without disorder. When the finite channel is introduced, we consider first a periodic system without disorder and then introduce disorder into the channel.
This allows us to probe the properties of the disordered chain as a function of the length and mimics typical experimental configurations, where the quantum system is weakly coupled to a mesoscopic conductor before eventually reaching the lead.

In the limit of weak TLS-to-chain coupling, we find that the decoherence time is given by $\ \tau_\phi\sim R$, where $R$ (resistance) is the inverse transmission coefficient of the lead evaluated at the eigenvalues of the TLS. This implies that $\tau_\phi$ undergoes mesoscopic fluctuations in analogy to mesoscopic conductors. The fluctuations are a function of the eigenenergies of the TLS, which depend on the TLS parameters, such as detuning and on-site energies.
	
Interestingly, the TLS can also be used as a dynamic probe for transport, since measuring the decoherence time is an indirect measurement of the resistance of the lead, via $\ \tau_\phi\sim R$. Hence we have identified a way to use a time-dependent measurement to obtain the static resistance of a disordered quantum wire. Note that the dependence is opposite to the usual resistance-fluctuation-induced dephasing, which scales as $\ \tau_\phi\sim R^{-1}$ \cite{pilgram2002efficiency}. The difference in scaling is due to thermal voltage fluctuations, which scale as $\de V\sim RT$, where $T$ is the temperature.

Finally, when the TLS is connected to a disordered lead the time dependence has recurring oscillations. These are similar to oscillations observed in some spin-based qubits in double quantum dots in the two electron spin blockade regime, where the electronic spins are coupled to a random nuclear spin bath \cite{ono2004nuclear}. These oscillations can also be observed in other regimes, beyond the two electrons spin blockade regime, when the double quantum dot is coupled to a nuclear spin bath \cite{harack2013power}. Our model can also be applied to spins coupled to a linear disordered spin chain using a Jordan-Wigner transformation \cite{lieb1961two,Rev1}.

The paper is organized as follows. In Section \ref{sec-review}, we briefly review two elementary systems, the isolated two-state system and the semi-infinite uniform chain, to remind the reader of some known results and to establish notation. In both cases, the Green's function is presented. In Section \ref{sec-2level+chain}, we couple these two systems, considering two cases: firstly treating the semi-infinite chain as a gate held at a controllable voltage, and secondly treating it as a lead at the same chemical potential as the two-state system. In Section \ref{sec-tripartite}, we consider a tripartite system, inserting a finite linear chain between the two-state system and the infinite lead. Again, two cases are considered: where the finite chain is ordered and where it is disordered. In the final section we summarize our findings and discuss avenues for further work.

\section{Isolated qubit and semi-infinite lead}
\label{sec-review}

\subsection{Isolated TLS}\label{subsec-2levelsys}
First we consider an isolated qubit, a TLS governed by the Hamiltonian
\be
H_{DD}=\left(\begin{matrix}
\eps_1 & \tau\\
\tau & \eps_2
\end{matrix}\right)
=\left(\begin{matrix}
\epsbar+\dezero/2 & \tau\\
\tau & \epsbar-\dezero/2
\end{matrix}\right),\label{eq-HDD}
\ee
where for the uncoupled system $\eps_i$ are the energies of the basis states $\ket{\eps_i}$, $\epsbar=(\eps_1+\eps_2)/2$ is the average energy and $\dezero=\eps_1-\eps_2$ is the energy splitting. The coupling $\tau$ is taken to be real and positive. It is convenient to also define the energy splitting and the energies of $H_{DD}$; these are $\de=\sqrt{\dezero^2+4\tau^2}$ and $\la_\pm=\epsbar\pm\dezero/2$. The corresponding Green's function in the uncoupled basis is given by
\be
\label{dog2}
G^{DD}(E)=\left(\begin{matrix}
E-\eps_1 & -\tau\\
-\tau & E-\eps_2
\end{matrix}\right)^{-1}.
\ee
The components are easily evaluated; for instance,
\ba
\label{G12E-isolated}
G^{DD}_{12}(E)
&=& \frac{\tau}{(E-\eps_1)(E-\eps_2)-\tau^2}
\nonumber\\
&=& \frac{\tau/\delta}{E-\la_+ + i0^+}-\frac{\tau/\delta}{E-\la_- + i0^+},
\ea
where the infinitesimal positive quantity $0^+$ gives the pole prescription necessary to compute the retarded time-dependent Green's function. This is obtained by Fourier transformation, giving zero for $t<0$ while for $t>0$ (with $\hbar=1$)
\ba
\label{G12t-isolated}
G^{DD}_{12}(t)&=&\int_{-\infty}^\infty dE\, e^{-iEt}G^{DD}_{12}(E)\nonumber\\
&=&-\frac{2\pi i\tau}{\de}\left(e^{-i\la_+t}-e^{-i\la_-t}\right)
\nonumber\\
&=&-\frac{4\pi \tau}{\de}e^{-i\epsbar t}\sin(\de t/2).
\ea
Similarly,
\be
\label{G11t-isolated}
G^{DD}_{11}(t)=\displaystyle-\frac{2\pi i}{\delta} e^{-i\epsbar t} \left( \delta\cos(\de t/2) - i\dezero\sin(\de t/2) \right).
\ee

In the context of qubit experiments, the most commonly measured quantity is the occupation probability of one of the states. If we consider a system in state $\ket{\eps_1}$ at $t=0$, then the probability of finding it in the state $\ket{\eps_2}$ at a later time $t$ is $P_{\eps_2}(t)=|G_{12}^{DD}(t)|^2/4\pi^2$. Since $\la_\pm$ are real, we get oscillatory behaviour, as expected. In most experiments on coherent oscillations in TLSs (spin qubits or charge qubits), $P_{\eps_2}(t)$ is the quantity most commonly measured. Hence, many of the graphs below display $|G_{12}^{DD}(t)|$.

When a qubit or TLS is coupled to a bath, one generally introduces the reduced density matrix $\rho^r$ of the TLS \cite{weiss1999quantum,leggett1987dynamics}. The decoherence rate is then associated with the decay of the off-diagonal element  $\rho^r_{12}(t)$ while the relaxation rate is given by the decay of the diagonal element $\rho^r_{11}(t)$. In the situation we consider here, all elements of $\rho^r$ decay with the same rate because we consider a weak tunneling coupling to the TLS. Hence, the decoherence rate, $\tau_\phi^{-1}$, the relaxation and population decay rates are all equal. Indeed,  $4\pi^2\rho^r_{11}=|G^{DD}_{11}|^2$ and $4\pi^2\rho^r_{12}=G^{DD}_{11}{G^{DD}_{12}}^*$, and as can be seen from (\ref{G12t-isolated},\ref{G11t-isolated}), the different components of $G^{DD}$ have the same decay rates, as do the different components of $\rho^r$.

In the remainder of this paper we will use the terminology of decoherence rate, $\tau_\phi^{-1}$, since it illustrates the coherent nature of the oscillations of the TLS.

\subsection{Semi-infinite chain}\label{subsec-semiinfchain}
Let us now consider a uniform semi-infinite chain described by the Hamiltonian
\be
H_\infty=\left(\begin{matrix}
0 & t_0 & 0 & \hdots\\
t_0 & 0 & t_0 &\hdots\\
0 & t_0 & 0 &\hdots\\
\vdots & \vdots & \vdots & \ddots
\end{matrix}\right),
\label{TBH}
\ee
with $t_0$ taken real and positive. In fact, to reduce clutter, throughout this paper we will take $t_0$ as the unit of energy by putting $t_0\to1$ in \eqref{TBH}. Energies and times are then dimensionless quantities, and to recuperate dimensionful quantities we must multiply (divide) energies (times) by $t_0$.

\begin{figure}[h!]
	\centering
	\includegraphics[width=0.5\textwidth]{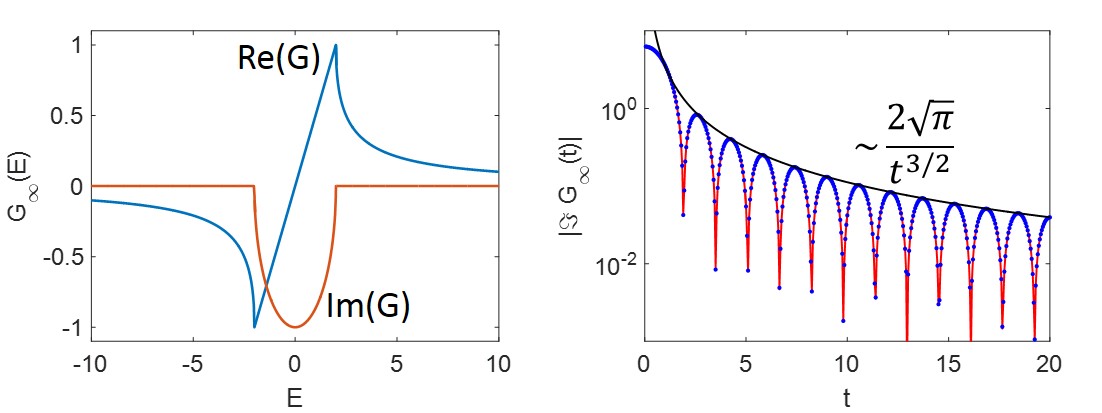}
	\caption{(color online) Left: the real and imaginary parts of the energy-dependent surface Green's function $G^S_\infty$. Right: the time dependence of the surface Green's function evaluated analytically (\eqref{GSofT}, blue), as well its asymptotic time dependence ($\sim t^{-3/2}$), and via numerical evaluation of the Fourier transform of \eqref{GSofE} (red). The excellent agreement between the analytical and numerical curves was achieved by integrating over a wide energy range ($|E|\le 1000$) in order to correctly capture the high-frequency oscillations shown on the right.
	 }
	\label{semiinfinitechain}
\end{figure}

The Green's function $G_\infty(E)=(E-H_\infty)^{-1}$ can be determined without difficulty (see, {\it e.g.}, \cite{d1990conductance,datta2005quantum}); we are mainly interested in the surface Green's function (that for the first site), which is
\be
G^S_\infty(E)=\frac{E-\sgn(E+2)\sqrt{E^2-4}}{2}.
\label{GSofE}
\ee
The square root is defined to have a positive imaginary part if its argument is negative, so
\be
G^S_\infty(E)=\frac{E-i\sqrt{4-E^2}}{2}\quad\mbox{if}\quad |E|<2.
\label{GSofE2}
\ee
The local density of states is given by $-\Im  (G^S_\infty)$ and outside the band $G^S_\infty\to 0$ as $E\to\pm\infty$ (see Fig.~\ref{semiinfinitechain}).

The time-dependent Green's function is the Fourier transform of \eqref{GSofE}:
\be
G^S_\infty(t)= FT \{G^S_\infty(E)\} = -\frac{2\pi i}{t} J_1(2t).
\label{GSofT}
\ee

This result is similar to early results on harmonic chains, such as the Rubin model \cite{rubin1963momentum}, where a heavy mass $M$ is coupled to a semi-infinite chain, where the damping kernel is given by
\be
\gamma=\frac{m\omega_L}{Mt} J_1(\omega_Lt).
\ee
which is equivalent to our \eqref{GSofT} above. Here $\omega_L$ is the maximum frequency. The very long time dependence will always be bound by this limiting behaviour, which determines how quickly a state can escape to infinity (quantum diffusion). In this work, however, we are more interested in the shorter time scales, where the decoherence of the TLS is changing exponentially with time. The exponential behaviour eventually saturates to a power law when reaching the quantum diffusion regime \cite{fiori2006non}.

\section{TLS coupled to semi-infinite chain}\label{sec-2level+chain}
Coupling the two systems via a hopping term is achieved with the Hamiltonian
\be
H=\left(\begin{array}{c|c}
H_{DD} & V\\
\hline
V^\dagger & H_\infty
\end{array}\right)
\label{2L&SIC}
\ee
with $H_{DD},\ H_\infty$ as above and
\be
V=\left(\begin{matrix}
0 & 0 & \cdots\\
\tc & 0 & \cdots
\end{matrix}\right)
\label{eq-V}
\ee
with $\tc$ taken real and positive.
Using standard methods, the Green's function for \eqref{2L&SIC} can be found, and in particular the effect of the semi-infinite chain on the double dot can be described elegantly by a modified Green's function for the double dot:
\be
\label{dog1}
G^{DD}_\infty(E)=\left(\begin{matrix}
E-\eps_1 & -\tau\\
-\tau & E-\eps_2-\Si_\infty(E)
\end{matrix}\right)^{-1}
\ee
where the self-energy is
\ba
\Si_\infty(E)&=& \tc^2 G_\infty^S(E)\nonumber\\
&=&\frac{\tc^2}{2}\left(E-i\sqrt{4-E^2}\right),
\label{eq-Sigma}
\ea
the latter expression being valid if $|E|<2$.

It is interesting and useful to note that \eqref{dog1} is just the Green's function of the isolated TLS \eqref{dog2} with the replacement $\eps_2 \to \eps_2+\Si_\infty(E) \equiv \eps_{2,\infty}(E)$. The individual components of the energy-dependent Green's function are then given by (\ref{G12E-isolated},\ref{G12t-isolated}) (and similar expressions for the other components) with this replacement as well as $\de \to \de_\infty(E)$, $\la_\pm \to \la_{\pm,\infty}(E)$, where $\de_\infty(E)$ is just $\de$ with $\eps_2 \to \eps_{2,\infty}(E)$ and similarly for $\la_{\pm,\infty}(E)$. However, the time-dependent Green's function cannot be determined from \eqref{G12t-isolated} by a similar replacement, since the energy dependence of the self-energy affects the Fourier transform in a highly nontrivial manner.

An effective Hamiltonian for the double dot can be read off \eqref{dog1}; it is simply \eqref{eq-HDD} with the replacement $\eps_2\to\eps_{2,\infty}(E)$. From \eqref{GSofE}, we see that $\Si_\infty(E)$ is complex if $|E|<2$, giving rise to a non-Hermitian effective Hamiltonian. This simply reflects the fact that current can flow between the double dot and the semi-infinite chain, so from the point of view of the double dot probability need not be conserved. Indeed, the time-dependent double dot Green's function shows exponential behaviour, as we shall now see.

\subsection{Coupling to a gate}\label{subsection-couplegate}
\begin{figure}[h!]
	\centering
	\includegraphics[width=0.5\textwidth]{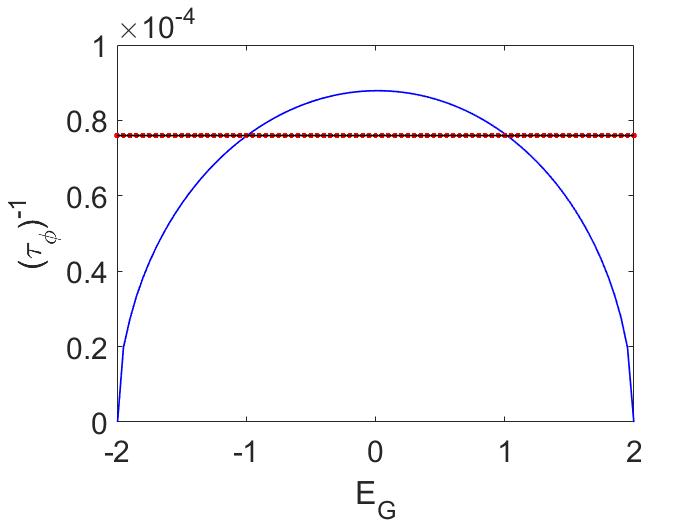}
	\caption{(color online) Blue line: Gate voltage dependence of the decay rate from \eqref{tau_phi-inverse2}. The horizontal lines (black: numerical determination of decay rate; red: approximate decay rate as given in \eqref{coupletolead}) are for the lead case, discussed in the next section. The excellent agreement between the two lines illustrates the validity of the approximation used to derive \eqref{coupletolead} (see Appendix \ref{app-eq15})
	Parameter values: $\eps_1=-\eps_2=1$, $\tc=0.1$, $\tau=0.19$.
}
	\label{fig-gate}
\end{figure}

For the case where the semi-infinite chain is a gate, we can model this using $\Si_\infty(E_G)$, where $E_G$ is the Fermi energy of the gate contact. This is a reasonable model for the case where the Fermi energy of the gate is not affected by the TLS. In this case, $E_G$ simply characterizes the properties of the gate, mainly via its local density of states (and assuming zero temperature throughout). In other words, we neglect the self-energy term of the TLS on the gate. This is quite different to the case where a lead is coupled to a TLS, where we need to include the self-energy term of the TLS on the gate, which we treat in the following section. This is equivalent to assuming that we shift the electrostatic energy of the TLS with respect to the gate potential. In this case $E_G$ is simply describing the gate parameter and we can use the results derived in (\ref{G12t-isolated}) directly with the substitution $\eps_2 \rightarrow \eps_{2,\infty}(E_G)$. Assuming $|E_G|<2$, $\la_{\pm,\infty}$ are complex, so the Green's function consists of damped oscillating functions. The slowest exponential determines the decoherence time scale; we find
\be
\label{tau_phi-inverse1}
({\tau_\phi})^{-1}=\min\left(-\frac{1}{2}\Im\left\{\Si_\infty(E_G)\pm\de_\infty(E_G)\right\}\right).
\ee
A more explicit expression can be given by defining $\varepsilon_G=\sqrt{4-E^2_G}$, in terms of which $\Im\left\{\Si_\infty(E_G)\right\}
=-\tc^2\varepsilon_G/2$.
Writing $\de_\infty(E_G)=\sqrt{a+ib}$ where
\be
\label{aandb}
\begin{array}{rcl}
a &=& \left( \dezero - \tc^2 E_G/2 \right)^2
- \left(\tc^2\varepsilon_G/2\right)^2 +4\tau^2,\\
	b &=&  2 \left( \dezero - \tc^2 E_G/2 \right)
	\tc^2\varepsilon_G/2,
\end{array}
\ee
we find
\be
\label{tau_phi-inverse2}
({\tau_\phi})^{-1} = \frac{\tc^2}{4}\varepsilon_G
- \left(\frac{\sqrt{a^2+b^2}-a}{8}\right)^{1/2}.
\ee
The decoherence rate varies with the gate energy, as shown in Fig.~\ref{fig-gate}.

If the coupling between the double dot and the chain is weak (that is, if $\tc^2\ll 1$), we can expand \eqref{tau_phi-inverse2}. We find
\be
\left( \tau _{\phi } \right)^{-1}\approx
\left( 1-\frac{|\dezero|}{\de} \right)\frac{\tc^2}{4}\varepsilon_G
\label{eq-tauphiinv-approx1}
\ee
with corrections of order $\tc^4$. Finally, note that if in addition $\tau\ll |\dezero|$, $\de\approx|\dezero|$ and there is an almost-complete cancellation between the two contributions to the decay rate, giving
\be
\left( \tau _{\phi } \right)^{-1}\approx
\frac{\tc^2\tau^2}{2\dezero^2}\varepsilon_G
\label{luna}
\ee
with corrections of order $\tc^4$ and $\tc^2\tau^2$.

\subsection{Coupling to a lead}\label{subsection-couplelead}
For the case where the semi-infinite chain is a lead, the situation is more complex since we must use the energy-dependent self-energy $\Si_\infty(E)$. Here we assume that the lead reservoir is empty, hence the TLS will eventually leak out into the empty reservoir. The exact result would consist in taking the Fourier transform of the components of \eqref{dog1}, which cannot be done analytically. However, if $\tc^2\ll 1$, an approximate analytical expression can be obtained by evaluating the energy-dependent quantities at the poles of the isolated TLS: $\Si_\infty(E)\to\Si_\infty(\la_\pm)$ and $\de_\infty(E)\to\de_\infty(\la_\pm)$. (The validity of this approximation is discussed in Appendix \ref{app-eq15}.) As with the gate, the slowest exponential determines the decoherence time scale, and we find
\be
\left(\tau_\phi\right)^{-1} \approx \min \left( -\frac{1}{2}\Im\left\{\Si_\infty(\la_\pm)
\pm\de_\infty(\la_\pm)\right\} \right)
\label{coupletolead}
\ee
which is just \eqref{tau_phi-inverse1} with $E_G\to\la_\pm$. This result is shown as the red horizontal line in Fig.~\ref{fig-gate}, where the parameter values chosen give $\la_\pm\approx\pm1.02$. Also illustrated (black horizontal line) is the decay rate obtained by studying the large-time behaviour of the numerically-evaluated Fourier transform of \eqref{dog1}. The excellent agreement between the two is a convincing justification of the approximation used in deriving \eqref{coupletolead} (see Appendix \ref{app-eq15}).

For other parameter values (maintaining $\tc^2\ll 1$), \eqref{coupletolead} can be expanded in powers of $\tc^2$, which yields
\ba
(\tau_\phi)^{-1}&\approx& \frac{\tc^2}{4} \min \left\{
\left( 1\mp\frac{\dezero}{\de} \right)\right.\nonumber\\
&&\left. \times\sqrt{4-\frac{\eps_1^2+\eps_2^2+2\tau^2\pm2\epsbar\de}{2}} \right\}
\label{eq-tauphiinv-approx2}
\ea
with corrections of order $\tc^4$.

For the special case $\eps_{1}=-\eps_{2}=\dezero/2$ (thus $\epsbar=0$), the decoherence rate can be further simplified to
\begin{equation}
\left( \tau_{\phi }\right) ^{-1} \approx  \frac{\tc^2}{4}\left(
1-\frac{|\dezero|}{\de }\right) \sqrt{%
4-(\dezero/2)^{2}-\tau^{2}},
\end{equation}
which is in excellent agreement with \eqref{tau_phi-inverse2} for the parameter values used in Fig.~\ref{fig-gate}.

When comparing the lead case to the gate case, it is important to note that the decoherence rate is maximal when the gate potential lies in the middle of the band. Interestingly, decoherence can be strongly suppressed near the band edges, even though the density of states in one dimension diverges there. Physically, the band edge describes the low-energy modes of a parabolic band. Hence, lowering the Fermi energy of the gate (or, equivalently, its charge density) leads to a substantially reduced decoherence rate. This is in contrast to the lead case, where the decoherence rate is mainly determined by the coupling strength.

This leads us to the next section, where we insert a finite chain between the double dot and the lead in order to map realistic systems. 


\section{Tripartite system}\label{sec-tripartite}
We now consider the case illustrated in Fig.~\ref{sketch}, where we insert a finite linear chain (of length $N$) between the double dot and the lead. Two cases are considered: where the finite chain is ordered and where it is disordered.

\subsection{Ordered finite chain}\label{subsec-ordered}
Suppose the system is described by the Hamiltonian
\be
H=\left(\begin{array}{c|c|c}
H_{DD} & V_N & 0\\
\hline
V_N^\dagger & H_N & W\\
\hline
0 & W^\dagger & H_\infty
\end{array}\right)
\label{H-dot+chain+lead}
\ee
where $H_N$ is an $N\times N$ truncation of \eqref{TBH} (with $t_0=1$), $V_N$ is a $2\times N$ truncation of \eqref{eq-V} and $W$ is an $N\times \infty$ matrix in the form of \eqref{eq-V} with the substitution $\tc\to \tL$ (assumed real and positive). Of course, if $\tL=1$, the finite chain joins seamlessly with the lead and we must recover the results obtained in Sec.~\ref{sec-2level+chain}. The behaviour of the system is more interesting when we vary $\tL$. In this case, as we shall see, the decoherence rate depends strongly on the length of the inserted chain.

In order to see this, we follow the procedure of Sec.~\ref{sec-2level+chain}, including the effect of the lead on the chain by adding a self-energy term to the chain Hamiltonian, so that
\be
H\to\left(\begin{array}{c|c}
H_{DD} & V_N\\
\hline
V_N^\dagger & H_{N\infty}
\end{array}\right)
\label{eq-dot+chain+self-energy}
\ee
where
\be
H_{N\infty}=\left(\begin{matrix}
	0 & 1 & 0 & \cdots & 0 \\
	1 & 0 & 1 &\ddots & \vdots \\
	0 & 1 & \ddots &\ddots & 0 \\
	\vdots & \ddots & \ddots & 0 & 1 \\
	0  & \cdots & 0 &1 & \Si'_\infty \\
\end{matrix}\right).
\label{eq-HNinfty}
\ee
The self-energy $\Si'_\infty$ is as in \eqref{eq-Sigma} with the replacement $\tc\to \tL$.

We can repeat this, capturing the effect of the finite chain including self-energy with a new self-energy added to the double-dot Hamiltonian. The Green's function for $H_{N\infty}$ is defined by
\[
G_{N\infty}=(E-H_{N\infty})^{-1};
\]
we show in Appendix \ref{sec-app2} that the surface Green's function (that for the first site) is
\ba
G^S_{N\infty}&=&
\frac{s_N - \Si'_\infty s_{N-1}}{s_{N+1} - \Si'_\infty s_N}\nonumber\\
&=&
\frac{s_N - \tL^2e^{-ik} s_{N-1}}{s_{N+1} - \tL^2e^{-ik} s_N}.
\label{eq-analytical}
\ea
Here $k\in[0,\pi]$ is defined by
\be
\label{definek}
2 e^{-ik}=E-i\sqrt{4-E^2}
\ee
and we have written $\sin(Nk)=s_N$, {\it etc}.

The self-energy describing the effect of the finite chain and semi-infinite lead on the double dot is
\be
\Sni=\tc^2 G^S_{N\infty}
=\tc^2
\frac{s_N - \tL^2e^{-ik} s_{N-1}}{s_{N+1} - \tL^2e^{-ik} s_N}
\label{eq-analytical2}
\ee
and the Green's function for the double dot including these effects is
\be
\label{dog3}
G^{DD}_{N\infty}(E)=\left(\begin{matrix}
E-\eps_1 & -\tau\\
-\tau & E-\eps_2-\Sni(E)
\end{matrix}\right)^{-1},
\ee
which is just  \eqref{dog1} with the replacement $\Si_{\infty}\to\Sni$. Thus, instead of \eqref{coupletolead}, the decoherence time is given by
\be
\left(\tau_\phi\right)^{-1} \approx \min \left( -\frac{1}{2}\Im\left\{\Sni(\la_\pm)
\pm\de_{N\infty}(\la_\pm)\right\} \right)
\label{coupletolead2}
\ee
where $\Sni(\la_\pm)$ is given by \eqref{eq-analytical2} evaluated at momenta $k_\pm$ which are defined by \eqref{definek} with $E\to\la_\pm$, and $\de_{N\infty}$ is $\de_\infty$ with the substitution $\Si_\infty\to\Sni$.

Appendix \ref{app-eq15} provides a justification of \eqref{coupletolead}, and it is argued there that this also applies to \eqref{coupletolead2}. However, as argued at the end of that appendix, the validity of the approximation used to derive \eqref{coupletolead2} must be considered carefully; the conclusion is that this equation is valid if $\tc^2 N\ll1$.

The simplicity of the substitution $\Si_{\infty}\to\Sni$ in the above equations belies the fact that this substitution leads to much richer behaviour, as we will see. However, the added complexity simplifies in two limits, as it must. First, as mentioned above, if $\tL=1$, the finite and infinite chains connect seamlessly, so $G^S_{N\infty}$ must reduce to $G^S_{\infty}$ (see \eqref{GSofE2}), and indeed it does, as straightforward algebra demonstrates. Second, if $N\to\infty$ the infinite chain decouples and $G^S_{N\infty}$ must again reduce to $G^S_{\infty}$. Since \eqref{eq-analytical} is a ratio of rapidly oscillating terms, the limit $N\to\infty$ is not obvious, but we can factor out a rapid oscillation (for example, $e^{iNk}$) from numerator and denominator and drop the rapidly-oscillating terms compared to those which are more slowly oscillating (an approximation known as the rotating-wave approximation in quantum optics); this procedure shows that indeed $\lim_{N\to\infty}G^S_{N\infty} = G^S_{\infty}$.

While in general $\Im\{\Sni(\la_+)\}\neq\Im\{\Sni(\la_-)\}$ and this must be taken into account when evaluating \eqref{coupletolead2}, the situation simplifies somewhat if $\eps_1=-\eps_2=\dezero/2$ (which we will assume in what follows) because then $\la_-=-\la_+$, $k_-=\pi-k_+$, and $\Im\{\Sni(\la_+)\}=\Im\{\Sni(\la_-)\}$.

The decoherence rate \eqref{coupletolead2} can be made more explicit by first noting that, from \eqref{eq-analytical2},
\be
\Im\{\Sni\} = -\frac{\tc^2\tL^2 s_1^3}{(s_{N+1} - \tL^2s_Nc_1)^2 + (\tL^2s_{N}s_1)^2}.
\label{panda1}
\ee

Secondly, if we take $|\dezero|\sim1$ and $t_c^2,\tau^2\ll1$, $\dni$ can be expanded in powers of $\Sni$ (or, equivalently, $t_c^2$) and $\tau^2$, giving
\be
\Im\{\dni\} \approx -\sgn(\dezero)\left(1-\frac{2\tau^2}{\dezero^2}\right)\Im\{\Sni\}
\label{delbar}
\ee
with corrections of order $t_c^4\tau^2$ and $t_c^2\tau^4$. If $\dezero$ is positive (negative), there is an almost-complete cancellation between the two terms in the upper (lower) sign of \eqref{coupletolead2}. Irrespective of the sign of $\dezero$, we find
\be
\label{panda2}
\left(\tau_\phi\right)^{-1} \approx \frac{\tau^2}{\dezero^2}
\frac{\tc^2\tL^2 s_1^3}{(s_{N+1} - \tL^2s_Nc_1)^2 + (\tL^2s_{N}s_1)^2}
\ee
where the trigonometric functions can be evaluated using either of the momenta $k_\pm$. (For definiteness, we will use $k_+$ below.)

\begin{figure}[h!]
	\centering
	\includegraphics[width=0.5\textwidth]{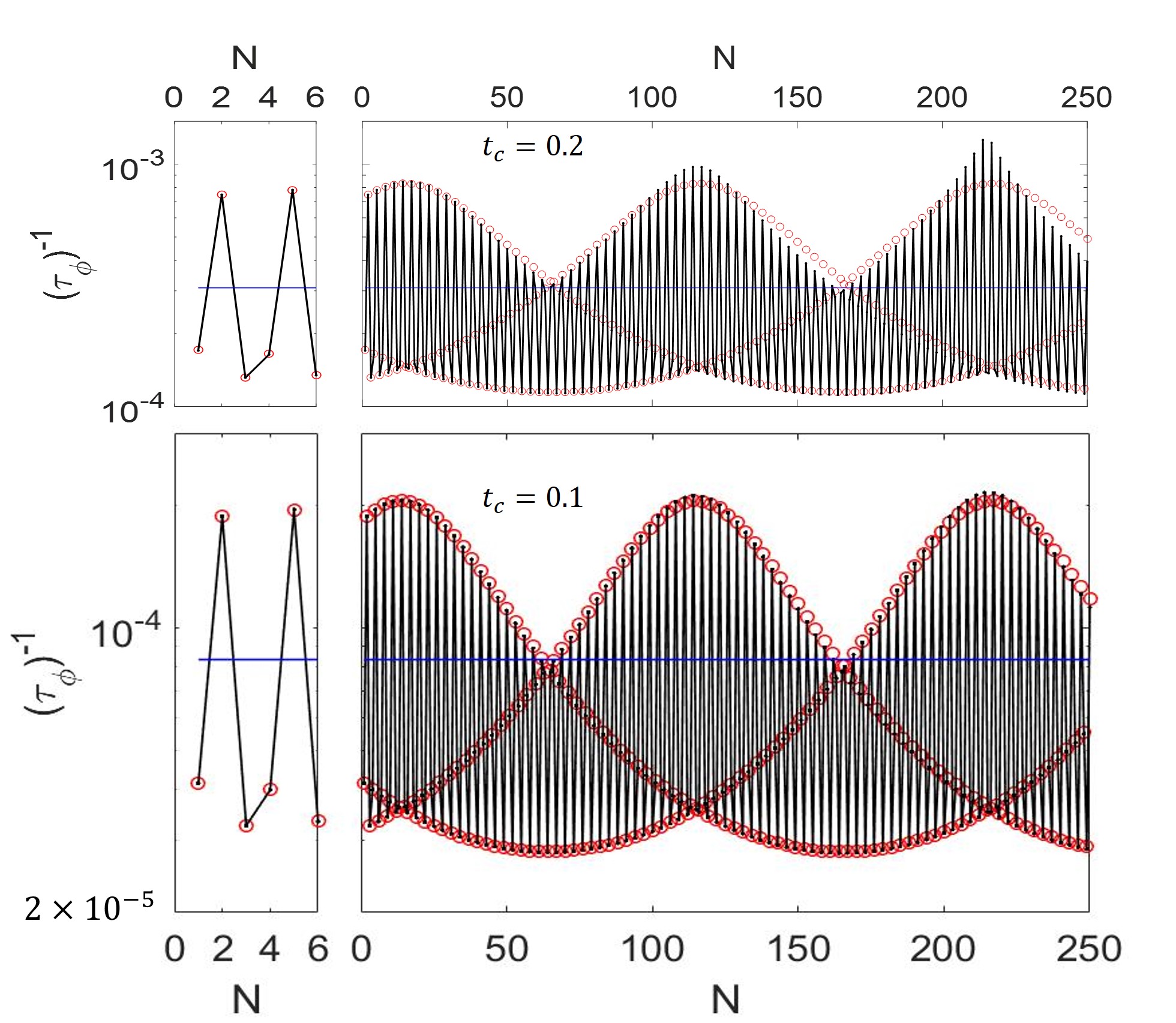}
	\caption{(color online)
$(\tau_\phi)^{-1}$ as a function of the chain length $N$ for $\eps_1=-\eps_2=1$, $\tau=0.19$, $\tc=0.1$ (bottom panels), $\tc=0.2$ (top panels), $\tL=0.65$. The left graph displays a small range of $N$ while the right side shows a wider range. The red circles are from the approximate expression \eqref{coupletolead2}, where the self-energy is evaluated at $\la_+$. The black dots and lines are obtained by evaluating $(\tau_\phi)^{-1}$ numerically using the decay of the time-dependent Green's function. The horizontal line is the rate obtained in the limit $N\rightarrow \infty$ (or equivalently, $\tL=1$).
Note that the analytical expression \eqref{coupletolead2} is only valid up to $\tc^2 N \sim 1$, as discussed at the end of Appendix \ref{app-eq15}. In fact, the agreement is still quite good throughout the lower panel, although a hint of growing disagreement can be seen (for instance, compare the black and red points at the three peaks). The upper panel shows greater disagreement between the two, as expected.}
\label{NdepAna}
\end{figure}

The decoherence rate $(\tau_\phi)^{-1}$ as a function of $N$ is displayed in Fig.~\ref{NdepAna} for representative values of the parameters (for which \eqref{panda2} and \eqref{coupletolead2} are in excellent agreement). The strong dependence on $N$ is particularly evident in the left panel which shows several small values of $N$. The larger range of $N$ in the right panel shows a very interesting pattern. In fact, the decoherence rate cycles between three values, each of which is slowly evolving, giving the appearance of three continuous curves which are translations of one another. The reason for this can be seen by examining \eqref{panda2}. Viewed as a continuous function of $N$, the decoherence rate has period $\pi/k_+ \approx 3.03$ for the parameter values used in the figure. But only integer values of $N$ are sampled in the figure; $N\to N+3$ produces only a miniscule variation in the decoherence rate, apparently slowing down the rate of change of the function in much the same way as the spokes of a moving wagon wheel appear slowed down or even reversed in a movie. Each of the three curves, corresponding to $N=3j,\ 3j+1,\ 3j+2\ (j\in\bbZ)$, is in fact a horizontally rescaled, translated version of the periodic behaviour of \eqref{panda2}. More precisely, if we define $f=(\pi/k_+-3)/3$ ($\approx0.01$ for the parameter values used in the figure), the curves are given by \eqref{panda2} with the replacements $N\to-Nf,\ 1-(N-1)f,\ 2-(N-2)f$. The period of these functions is $1/f$ larger than that of $(\tau_\phi)^{-1}$ ($\de N \approx 303$ for the parameter values used in Fig.~\ref{NdepAna}).

\begin{figure}[h!]
	\centering
		\includegraphics[width=0.25\textwidth]{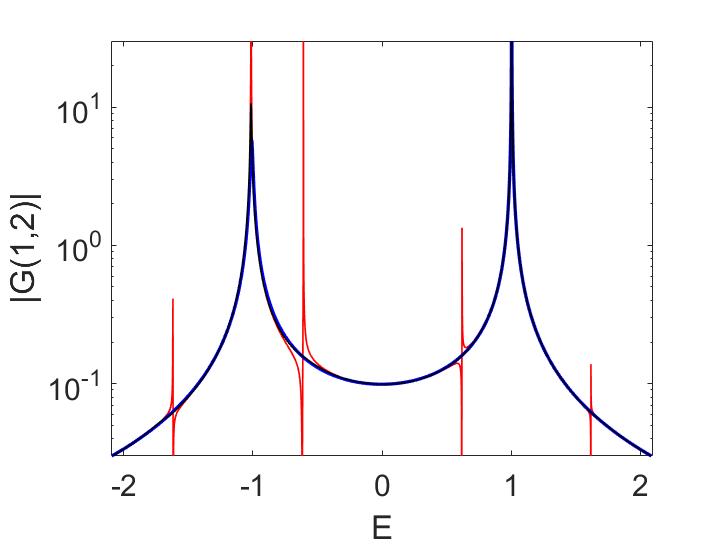}\includegraphics[width=0.25\textwidth]{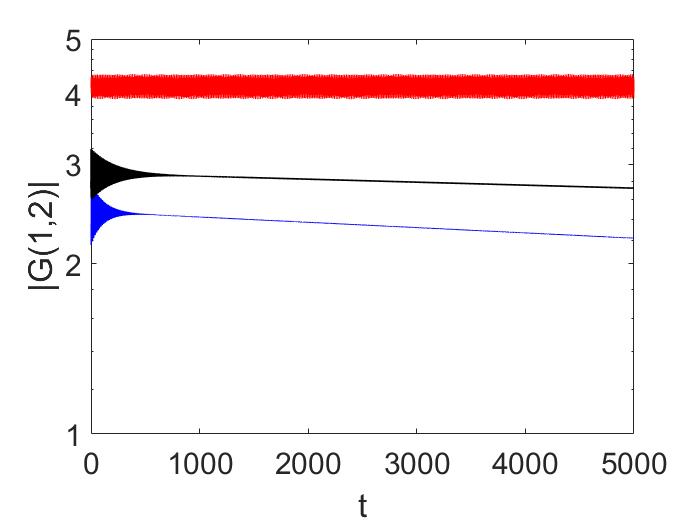}\\
	\includegraphics[width=0.25\textwidth]{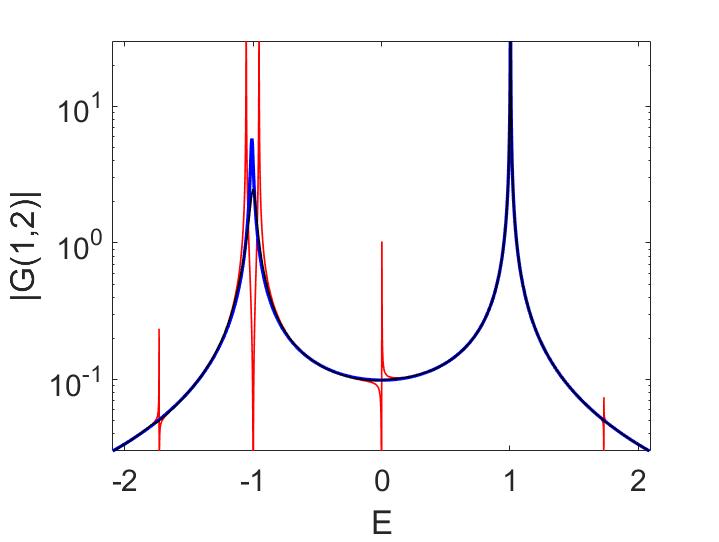}\includegraphics[width=0.25\textwidth]{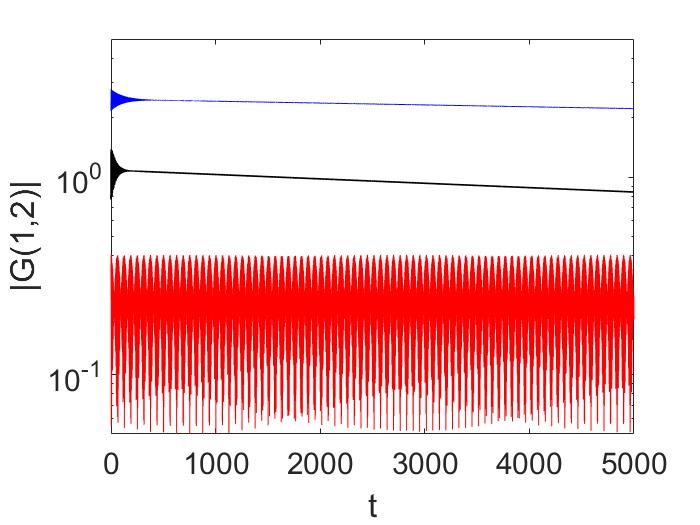}
	\caption{(color online) behaviour of Green's functions for $N=4$ (top, off resonance) and $N=5$ (bottom, on resonance) for the same parameter values as in the bottom panels of Fig.~\ref{NdepAna}. On the left is the energy dependence and on the right is the time dependence. The black lines correspond to the full system (double dot-segment-lead), as given in \eqref{dog3}, the red lines correspond to the dot connected to the segment without a lead (\eqref{dog3} with $\tL=0$), while the blue line corresponds to connecting the double dot directly to the lead (\eqref{dog3} with $\tL=1$). Note the increased decoherence rate in the resonant case.}
	\label{N3}
\end{figure}

Resonances occur when the eigenvalues of the chain match closely those of the double dot, leading to a greatly enhanced decoherence. Changing the dot energies changes the oscillation frequency as a function of $N$. We show in Fig.~\ref{N3} the details of what occurs for certain values of $N$.

The decoherence rate can be determined by taking the Fourier transform of the energy-dependent Green's function. We can use the approximate expression for the decoherence rate (\ref{coupletolead}) to compare the numerical Fourier transform and the analytical expression, as illustrated for representative parameter values in Fig.~\ref{NdepAna}.

The main conclusion here is that inserting a linear chain between the double dot and the lead leads to a non-monotonic dependence of the decoherence time. There are resonances whenever the energies of the chain match the double dot energies. This leads to a broadening of the energy dependence peak, which in turn yields an increased decoherence rate. Away from these resonances, the decoherence rate is suppressed and can provide an optimization scheme for decreasing the decoherence rate.

\subsection{Disordered finite chain}\label{subsec-disordered}

In the previous sections we only considered the clean case, where we assumed that the channel coupled to the TLS is clean without any defects. We now introduce defects and disorder into the channel as a way to model realistic experimental systems. We introduce the disorder by adding uniform uncorrelated random on-site potentials between $-va$ and $va$ on each lattice site of the chain elements (diagonal disorder). Adding disorder to the channel typically reduces the transmission probability due to enhanced scattering, eventually giving rise to exponential suppression with distance due to coherent interference (Anderson localization) \cite{anderson1958absence,gertsenshtein1959waveguides}. The transmission will vanish for strong disorder and hence decouple from the TLS. We will only consider disorder in the finite chain element, where we can change the length. This mimics the situation where the TLS is coupled to a disordered mesoscopic conductor which is then connected to a macroscopic contact. We start by evaluating the time dependence of the off-diagonal Green's function of the TLS as was done in the previous sections. We observe an overall exponential decay, from which we can extract an average decay rate or decoherence rate  $(\tau_\phi)^{-1}$.
\begin{figure}[h!]
	\centering
	\includegraphics[width=.5\textwidth]{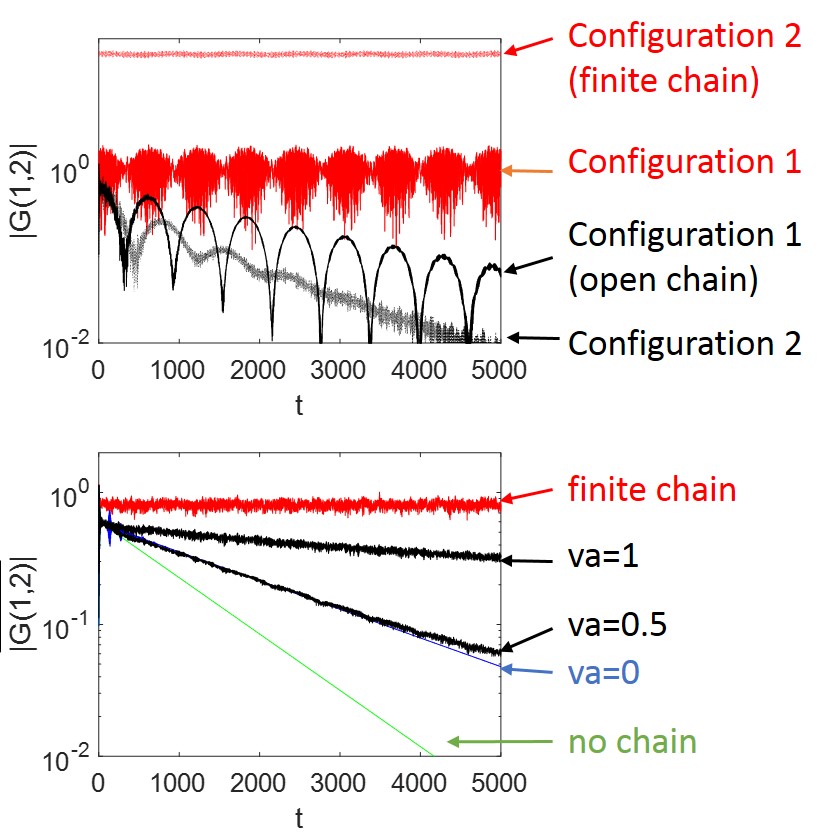}
	\caption{(color online) Top: off-diagonal element of the $DD$ Green's function as function of time for $va=0.5$ for two different disorder configurations 1 and 2. The parameter values used are the same as in Fig.~\ref{NdepAna}, except for $\tc=0.35$. The red traces correspond to having no semi-infinite lead attached to the end of the linear chain ($\tL=0$), while the black lines correspond to $\tL=0.65$. Bottom: time dependence of the configurational log-averages $\overline{|G_{1,2}|}\equiv e^{\vev{\log{|G_{1,2}|}})}$ for different values of the disorder strength ($va$=0, 0.5 and 1). We used 100 different configurations for the disorder average. The green trace corresponds to the case where no finite chain is placed between the semi-infinite lead and the TLS.}
	\label{DisorderTimeDep}
\end{figure}

As expected, the decoherence rate decreases with increasing disorder as illustrated in Fig.~\ref{DisorderTimeDep}. However, we also observe oscillations in the time dependence for certain disorder configurations with large periods. These oscillations average out when performing a configurational average. In many experimental systems, static disorder is not averaged out; hence, it is expected that some devices would show these oscillations. Since our linear chain can be mapped onto a spin chain, with neighboring spin-spin coupling, these oscillations have possibly the same origin as those observed in spin qubits coupled to a spin environment (nuclear bath) \cite{ono2004nuclear,harack2013power}. These oscillations stem from the build-up of a coherent superposition after dominant multiple-scattering events. This can lead to a characteristic period of oscillation of the envelope which is much longer than the intrinsic TLS oscillation period, as seen in Fig.~\ref{DisorderTimeDep}. Each different disorder configuration leads to a different resonance condition. 

\begin{figure}[h!]
	\centering
	\includegraphics[width=.5\textwidth]{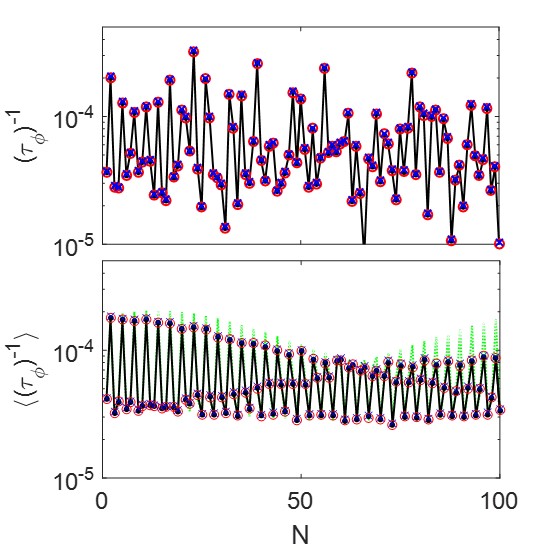}
	\caption{(color online) $(\tau_\phi)^{-1}$ as a function of the length of the disordered segment. The black dots and lines are obtained by evaluating $(\tau_\phi)^{-1}$ numerically using the decay of the time dependent Green's function for disorder strength $va=0.25$. All other parameter values are as in
the bottom panels of
Fig.~\ref{NdepAna}. The red circles are obtained by evaluating numerically the approximate expression (\ref{coupletolead}) with the self-energy evaluated at $\la_+$. The blue crosses are obtained from the transmission probability for the disordered segment evaluated numerically at $\la_+$ using equation \eqref{transmission}. Clearly, the agreement between the three methods is excellent. The upper panel is for a single disorder configuration whereas in the bottom panel we display the corresponding ensemble-averaged rate $\vev{(\tau_\phi)^{-1}}$, obtained by averaging over 100 different disorder configurations. For comparison purposes, we also show the non-disordered case ($va=0$) in the dotted green line of the lower panel.}
	\label{v025rand}
\end{figure}

\begin{figure}[h!]
	\centering
	\includegraphics[width=.5\textwidth]{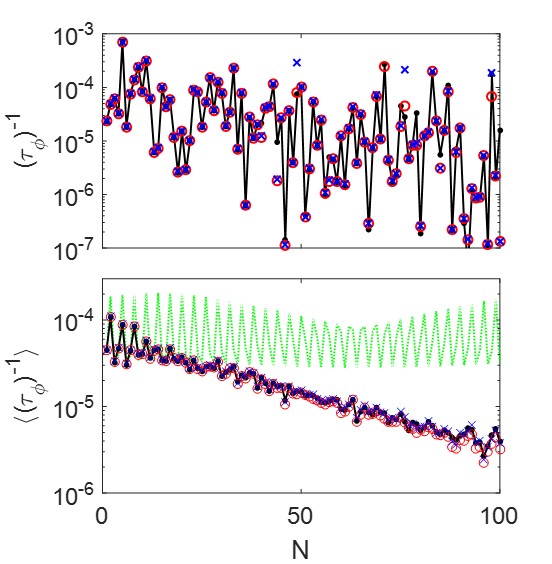}
	\caption{(color online) $(\tau_\phi)^{-1}$ as a function of the length of the disordered segment using the same parameter values and labels as in Fig.~\ref{v025rand}, except for a disorder strength of $va=1$. Here the disorder average was performed over 180 configurations.}
	\label{v1randb}
\end{figure}

Using equations \eqref{coupletolead2} and \eqref{delbar}, we can write 
\ba
\left(\tau_\phi\right)^{-1} &\approx&  -\frac{1}{2}\Im\left\{\Sni(\la_+)
+\de_\infty(\la_+)\right\}\nonumber\\
&\approx & -\frac{\tau^2}{\dezero^2} \Im\{\Sni(\la_+\})\nonumber\\
&\approx & -\frac{\tau^2}{\dezero^2}\frac{T(\la_+)}{4\Im \{G_\infty^S(\la_+)\} }
\label{transmission}
\ea
where we have used the relation between $\Sni(\la_+)$ and the transmission through the chain as detailed in Appendix \ref{sec-app3} (see \eqref{eq-koala}). This expression, which is one of our main results, is an excellent approximation as illustrated in Figs.~\ref{v025rand} and \ref{v1randb}. Hence, knowledge of the transmission evaluated at the TLS eigenvalue determines the decoherence properties. Expressed in terms of the two terminal resistance $R=1/T$, we therefore have $\tau_\phi\sim R$, which directly connects the dynamics of the TLS to the resistance of the disordered chain.

The decoherence rate $(\tau_\phi)^{-1}$ can be extracted by fitting the exponential decay, which can be done for different values of the chain length $N$. This leads to a dependence similar to that seen in the previous section (see Fig.~\ref{N3}), except that in the presence of disorder the rate is no longer a quasi-periodic function of $N$. Instead, for a given disorder configuration we observe strong fluctuations of the rate, as shown in Fig.~\ref{v025rand}. This is reminiscent of mesoscopic conductance fluctuations in finite coherent conductors \cite{lee1985universal,altshuler1985fluctuations,abrahams1980resistance}. The transmission can be described via a probability distribution and scaling behaviour \cite{anderson1980new,perel1984probability,pichard1986one}. In one dimension these fluctuations grow faster than the mean, due to the log-normal distribution of the transmission \cite{gertsenshtein1959waveguides,kramer1993localization}. For the tight binding model considered here, the average of the logarithm of the transmission is given by the Lyapounov exponent $\lambda_T\approx (1/2)\si^2/(4-E^2)$. This result is the lowest order in disorder strength, where $\si$ is the standard deviation of the on-site disorder and we have used a bandwidth of 2 ($t_0=1$) \cite{thouless1979ill}. Beyond the inverse Lyapounov exponent (localization length), the transmission is statistically zero. However, as discussed above, the transmission is characterized by strong fluctuations, which can be averaged out by averaging over different disorder configurations (ensemble average) as shown in Fig.~\ref{v025rand}. However, this depends on the disorder distribution. Certain disorder distributions can lead to ensemble-averaged fluctuations \cite{hilke2008ensemble}. Here we used a uniform disorder distribution and we can observe how the average rate converges towards quasi-periodic oscillations as a function $N$, while slowly decreasing with $N$. 

In general, only the configurational average of the transmission $\vev{T}$ can be computed with an analytical expression (assuming small disorder). Indeed, Pendry showed that for the disordered tight binding model $\log\vev{T}\approx-\lambda_T N/2$, while $\vev{\log(T)}\approx-\lambda_TN$, due to the log-normal distribution of the transmission \cite{pendry19821d,pendry1994symmetry}. These results are valid in the limit $N\rightarrow \infty$. For finite $N$,  $\vev{T}$ will oscillate with $N$ as shown in Figs.~\ref{v025rand} and \ref{v1randb} before reaching the limiting behaviour when $N\rightarrow \infty$ given by the Thouless Lyapounov exponent \cite{thouless1979ill}. For finite $N$ we are not aware of an analytic expression for $\vev{T}$ as a function of $N$. Therefore, for a single qubit coupled to a disordered chain, fluctuations in $(\tau_\phi)^{-1}$ will dominate and will make it very difficult to predict the decoherence time. However, if we consider a large number of qubits coupled to different disordered leads, the average decoherence rate $\vev{(\tau_\phi)^{-1}}$ is proportional to $\vev{T}$ (see \eqref{transmission}), but there will be large fluctuations of the rates between different qubits.

When increasing the disorder strength, as shown in Fig.~\ref{v1randb}, we observe a similar behaviour as for weak disorder except that the average rate tends exponentially to zero at a faster rate, which is a direct consequence of Anderson localization, since Anderson localization is characterized by the exponential decay of the transmission as a function of system length \cite{anderson1958absence}. Hence, the decoherence rate of the TLS is a one-to-one probe of Anderson localization. Conversely, if we want to decrease the decoherence rate of qubits, we have developed a simple scheme to do so effectively. Any lead or gate attached to the qubit needs to minimize the transmission at the eigenvalues of the TLS. Only the transmission at these eigenvalues is of any consequence. For instance, for continuous disordered systems, the transmission is minimized when the wavelength of the corresponding energy is comparable to the correlation length of the disorder potential \cite{eleuch2015localization}, which provides a natural guide to reducing the decoherence rate. An interesting application of using a TLS to probe localization is in the context of many-body localization (MBL), where finite 1D systems are usually considered \cite{basko2006problem}. In \cite{PhysRevB.91.140202,van2016single} it was suggested to use a coupled spin in order to probe these MBL states.

\section{Conclusions}\label{sec-conclusions}

In this paper, a tripartite system consisting of a qubit (TLS), a finite linear channel and a semi-infinite chain was considered. We showed that the decoherence rate of the qubit is given by the properties of the chain. In particular, the length of the linear channel leads to resonances in the decoherence rate, which we computed analytically. Fundamentally, the decoherence rate is shown to depend linearly on the transmission of the linear channel evaluated at one of the eigenvalues of the TLS. Adding disorder to the linear chain reduces the decoherence rate on average, which can be used in order to improve the performance of qubits coupled to leads or electronic gates. Since an electronic system can be mapped onto a spin chain, our results also apply to a couple of spins coupled to a disordered spin chain, where we observed oscillations with long periods similar to experimental spin qubits coupled to a disordered spin bath \cite{ono2004nuclear}. Finally, while it is perhaps more common to measure the decoherence rate by measuring the transport properties of the lead, the situation can be reversed: measurement of the decoherence rate of the TLS can be used to determine the resistance of the lead.

An interesting endeavor is to go beyond a single qubit and to consider multiple qubits coupled to a disordered channel. When qubits are coupled to the same infinite bath, this can lead to a much stronger decoherence rate (superdecoherence) \cite{palma1996quantum,reina2002decoherence}. However, for most cases relevant to experimental implementations, the scaling remains linear with the number of qubits \cite{ischi2005decoherence}. The exact manner in which disorder impacts the decoherence rate of multiple qubits would be interesting to investigate. Moreover, in this work we only considered the effect of relaxation on the decoherence rate. Considering pure dephasing as mechanism is expected to yield similar results when coupled to a disordered channel. 
	
\section*{Acknowledgments}
This work was supported in part by the Natural Science and Engineering Research Council of Canada and by the Fonds de Recherche Nature et Technologies du Qu{\'e}bec via the INTRIQ strategic cluster grant.

\appendix

\section{Justification of (\ref{coupletolead})}\label{app-eq15}
As explained in Section \ref{subsection-couplelead}, the energy-dependent Green's function of the double dot attached to a semi-infinite lead is obtained by the substitution $\eps_2\to\eps_{2,\infty}(E) = \eps_2 + \Si_\infty(E)$ with
\ba
\Si_\infty(E) &=& \frac{\tc^2}{2}\left( E-\sgn(E+2)\sqrt{E^2-4} \right)\nonumber\\
&=&\frac{\tc^2}{2}\left( E-i\sqrt{4-E^2} \right),
\label{eq-bird}
\ea
where in the last line we have assumed $|E|<2$. Thus, for instance, \eqref{G12E-isolated} becomes
\ba
G^{DD}_{12}(E)&=&
\frac{\tau}{\de_\infty(E)}\left(\frac{1}{E-\la_{+,\infty}(E) + i0^+}\right.\nonumber\\
&&\qquad\quad-\left.\frac{1}{E-\la_{-,\infty}(E) + i0^+}\right),
\label{GcDD12-1}
\ea
where
\ba
\de_\infty(E) &=& \sqrt{(\dezero-\Si_{\infty}(E))^2+4\tau^2},\\
\qquad
\la_{\pm,\infty}(E) &=& \epsbar+\frac{1}{2}\left(\Si_{\infty}(E)\pm\de_\infty(E) \right).
\ea
We are interested in the Fourier transform of \eqref{GcDD12-1}. Let us examine the first term (the second being similar). It is
\be
I \equiv \tau \int_{-\infty}^\infty dE \frac{e^{-iEt}}
{\de_\infty(E) \left(E-\la_{+,\infty}(E) +i0^+\right)}.
\label{eq-defI}
\ee
Suppose that $\eps_1,\eps_2,\tau$ are all $O(1)$ and that $\tc\ll 1$. Then $\de_\infty(E) \approx \de$ and $\la_{\pm,\infty}(E) \approx \la_\pm$ where $\de,\la_\pm$ are given in Section \ref{subsec-2levelsys}. The first factor in the denominator of \eqref{eq-defI} is never zero and will not affect the frequency of the Fourier transform. It is the second term which has a direct effect on the frequency. Defining $\si\equiv \tc^2/2$, if $\si=0$ then $\la_{+,\infty}(E)=\la_+$, giving a simple pole at $E=\la_+$ and frequency $\la_+$ as in \eqref{G12t-isolated}. If $\si$ is small (but nonzero) then presumably there is a zero of the second term near $\la_+$ which will dominate the integral. Let this zero be $E^*$, and suppose it is a simple pole. Then a direct application of the residue theorem implies
\[
I=({\rm const.})\,e^{-iE^*t}.
\]
Let us find $E^*$ perturbatively in $\si$. It is the solution of $E-\la_{+,\infty}(E)=0$. Writing $\la_{+,\infty}(E)=\la_+ + \si g(E) + O(\si^2)$, it is easy to show
\be
E^*= \la_{+,\infty}(\la_+) + \si^2 g(\la_+)g'(\la_+) + \cdots.
\label{eq-estar}
\ee
The function $g(E)$ is easily calculated (it is in fact $G^S_\infty(E)$ given in \eqref{GSofE} multiplied by a constant of order 1 for generic parameter values) but it is not necessary to do so; we need only note that $g(E)$ and its derivatives are $O(1)$.

If we drop terms beyond linear in $\si$, we have
\be
I=({\rm const.})e^{-i \la_{+,\infty}(\la_+) t}.
\label{eq-gerbil}
\ee
This analysis applied to the second term of \eqref{GcDD12-1} gives the same conclusion with $\la_{+,\infty}(\la_+)\to\la_{-,\infty}(\la_-)$. The decay rate of the Green's function is therefore determined by $\Im\{\la_{\pm,\infty}(\la_\pm)\}$. Given the definition of $\la_{\pm,\infty}(E)$, this completes the demonstration of the validity of \eqref{coupletolead}.

In Section \ref{subsec-ordered} we apply the same argument to the tripartite system with ordered finite chain, leading to \eqref{coupletolead2}, which is the that system's equivalent of \eqref{coupletolead}. However, the domain of validity of the approximation applied to the tripartite system merits discussion. For the bipartite system, in neglecting the $O(\si^2)$ term in \eqref{eq-estar} to arrive at \eqref{eq-gerbil}, we are neglecting a term of $O(\si^2)$ compared to one of $O(\si)$, so this is a reasonable approximation if $\si\ll1$, that is, if $\tc^2\ll1$. For the tripartite system, the function $g(E)$ is once again a constant of order 1 times the relevant Green's function, now $G^S_{N\infty}(E)$ given in \eqref{eq-analytical}. The latter is $O(1)$ but it varies on an $N$-dependent time scale; in particular, its derivative is $O(N)$. Thus, the final term shown in \eqref{eq-estar} is $O(\si^2N)$, and neglecting it compared to the $O(\si)$ term is reasonable if $\si N\ll1$, that is, if $\tc^2 N \ll 1$. Thus, even if $\si$ is tiny, for sufficiently large $N$ the approximation is not valid.

\section{Derivation of (\ref{eq-analytical})}
\label{sec-app2}
The Green's function $G_{N\infty}$ is defined by the $N\times N$ matrix equation
\be
G_{N\infty}(E-H_{N\infty})=\id
\label{eq-matrixequation}
\ee
with $H_{N\infty}$ given in \eqref{eq-HNinfty}. In order to streamline the equations somewhat, we define a ``vector" representing the first row of the Green's function: $g_j \equiv (G_{N\infty})_{1j}$ (the object of interest being $G^S_{N\infty}=g_1$). Then the first row of \eqref{eq-matrixequation} is the following set of equations:
\ba
E g_1 -  g_2 &=&1,\label{veqs-1}\\
- g_{j-1} + E g_j -  g_{j+1} &=& 0\quad (1<j<N),\label{veqs-j}\\
- g_{N-1} + (E-\Si'_\infty) g_N &=& 0.\label{veqs-N}
\ea
These equations can be most conveniently solved in terms of the momentum $k\in[0,\pi]$ defined by $2 e^{-ik}=E-i\sqrt{4-E^2}$. The solution of the middle equations \eqref{veqs-j} is
\[
g_j = A e^{ijk} + B e^{-ijk}
\]
where the coefficients $A$ and $B$ are determined by the boundary equations (\ref{veqs-1},\ref{veqs-N}). These can be written
\[
\left(\begin{array}{cc}u&v\\w&x\end{array}\right)
\left(\begin{array}{c}A\\B\end{array}\right)=
\left(\begin{array}{c}1\\0\end{array}\right)
\]
where
\ba
u&=&v=1,\nonumber\\
w&=&(e^{ik}-\Si'_\infty)e^{iNk},\nonumber\\
x&=&(e^{-ik}-\Si'_\infty)e^{-iNk}.\nonumber
\ea
The solution is straightforward and with some algebra we find
\[
G^S_{N\infty}=
\frac{s_N - \Si'_\infty s_{N-1}}{s_{N+1} - \Si'_\infty s_N}
\]
where we have written $\sin(Nk)=s_N$, {\it etc}. Note that $\Si'_\infty$ is complex and depends on $E$ (or on $k$); we can write $\Si'_\infty=\tL^2e^{-ik}$ which leads to the alternative expression
\[
G^S_{N\infty}=
\frac{s_N - \tL^2e^{-ik} s_{N-1}}{s_{N+1} - \tL^2e^{-ik} s_N},
\]
completing the demonstration of \eqref{eq-analytical}.

\section{Transmission derivation}
\label{sec-app3}

Here we evaluate the transmission coefficient of a linear chain of length $N$ between two semi-infinite leads characterized by their surface Green's function $G_\infty^S$ as discussed in \eqref{GSofE}. Assuming respective couplings $\tc$ and $\tL$ between the linear chain and the left and right semi-infinite leads, the transmission coefficient is given by

\be
T=4(\Im \{G_\infty^S\})^2\tL^2\tc^2|G_{1N}|^2,
\ee
where $G_{1N}$ is the off-diagonal element of the Green's function connecting the first site to the last site. The first diagonal element $G_{11}$ of the finite chain can be approximated to leading order in $\tc$ and $\tL$ by
\ba
G_{11}&\approx & G_{11}^0+|G_{11}^0|^2 \tc^2 G_\infty^S+|G_{1N}^0|^2 \tL^2 G_\infty^S\\
&\approx & G_{11}^0+|G_{1N}^L|^2 \tL^2 G_\infty^S.
\ea
We used $G_{11}^0$ for the first diagonal element when the chain is not coupled to any leads, while $G_{11}^L$ is the same element when the chain is only coupled to the left lead ($\tc$). Since $G_{11}^0$ is real inside the band, we have
\be
\Im\{ G_{11} \} \approx |G_{1N}^L|^2 \tL^2 \Im \{ G_\infty^S \}.
\ee
Finally, to leading order in chain-leads couplings we can write
\be
T\approx 4\Im G_\infty^S\tc^2\Im \{G_{11} \} = 4\Im\{ G_\infty^S \}\Im\{ \Sni\}
\label{eq-koala}
\ee
for the transmission coefficient, since $G_{11}$ corresponds to $G_{N\infty}^S$ in \eqref{eq-analytical}.

\bibliographystyle{apsrev4-1}
\bibliography{ref2}

\end{document}